\documentclass[preprint2,10.5pt]{aastex}
\begin{document}

\title{\large{\rm{NEW RESULTS FOR THE OPEN CLUSTER BICA 6 \\ AND ITS ASSOCIATED PLANETARY NEBULA ABELL 8}}}

\author{D.~G. Turner$^{1,5,}$\altaffilmark{*}, J.~M. Rosvick$^{2,}$\altaffilmark{*}, D.~D. Balam$^{3}$, A.~A. Henden$^4$, D.~J. Majaess$^{1,5}$, D.~J. Lane$^{1,5}$}
\affil{$^1$ Saint Mary's University, Halifax, NS, Canada}
\affil{$^2$ Thompson Rivers University, Kamloops, BC, Canada}
\affil{$^3$ Dominion Astrophysical Observatory, Victoria, BC, Canada}
\affil{$^4$ AAVSO, Cambridge, MA, USA}
\affil{$^5$ Abbey Ridge Observatory, Stillwater Lake, NS, Canada}
\email{\rm{turner@ap.smu.ca}}
\altaffiltext{*}{Visiting astronomer, Dominion Astrophysical Observatory, NRC-HIA, Victoria, BC, Canada}

\begin{abstract}
The likely membership of the planetary nebula Abell 8 (PN G167.0--00.9) in the open cluster Bica 6 is confirmed by CCD spectra, {\it UBV(RI)$_C$} photometry, and radial velocities for luminous cluster stars. The reddening, estimated distance, and radial velocity of the planetary nebula agree with parameters derived for Bica 6 of $E_{B-V}{\rm(mean)} = 0.42$, $d = 1.60 \pm0.11$ kpc, and $V_R = 57 \pm1$ km s$^{-1}$, with a cluster age of $10^9$ yr, a diagnostic blue hook, and a few blue stragglers, including a peculiar B1~Vnn star (HDE~277593) that may be a post-AGB star. The results identify Bica 6 as a potential calibrator of the planetary nebula distance scale. The central star of the planetary nebula has a reddening of $E_{B-V}=0.49\pm0.02$, with a possible circumnebular excess, and an estimated luminosity of $M_V = +7.44\pm0.16$. It is also an optical double in 2MASS images, with a likely progenitor according to evolutionary considerations being a late B-type dwarf of $\sim 2.3\,M_{\sun}$.
\end{abstract}

\keywords{techniques: spectroscopic---Galaxy: open clusters and associations: individual: Bica 6---ISM: planetary nebulae: individual: PN G167.0--00.9.}

\section{{\rm \footnotesize INTRODUCTION}}
The progenitors of planetary nebulae (PNe) are generally considered to have once been main-sequence stars with masses of up to $\sim 8\,M_{\sun}$ \citep[e.g.,][]{we00}, yet a practical demonstration of such a link has so far been limited by the paucity of PNe associated with intermediate-age open clusters, where physically-associated, lower-mass, main-sequence progenitors may exist. Searches for possible planetary nebula (PN)/open cluster associations have revealed a number of potential spatial coincidences, as noted by \citet{zi75} and Lubos Kohoutek \citep*[see][]{ma07}, but available data on possible matches in radial velocity, reddening, and distance of the PNe and cluster stars are as yet unconvincing \citep[e.g.,][]{ki08}. The relatively short duration of the PN evolutionary phase and the rapid rate of dissolution of open clusters conspire to make true physical links relatively rare, if they exist at all \citep{ma07}.

The distances to individual PNe are also difficult to establish, resulting in a variety of techniques applied to the problem \citep{ka85,lu89}. If the central star is bright enough for photometric or spectroscopic observation, the distance can often be established through trigonometric or spectroscopic parallax. Trigonometric parallaxes, for example, have been derived for several PNe through the U.S. Naval Observatory program \citep{pi93,ha97,ha07}, from measurements by the Hipparcos satellite \citep[see][]{ac98}, and most recently by measurements with the Hubble Space Telescope \citep[HST,][]{be09}. Spectroscopic parallaxes produce mixed results \citep[e.g.,][]{ac98}, which may relate to the models used and the quality of spectra for faint stars. Nebular expansion parallaxes have also been applied \citep{te97}, despite their observational limitations \citep{te97} and initial assumptions \citep{ka85}. Indirect methods also exist, the most basic being the Shklovsky distance method \citep{sh56}, which is tied to estimates for the surface brightness of the surrounding PN shell, thus being limited by the difficulties of recording the spectra of PNe. Improved versions of the technique \citep[e.g.,][]{zh95,ph04} typically involve measurement of the radio continuum surface brightness temperature $T_b$. Finally, it is possible to infer distances to PNe by using the reddening inferred from a comparison of line strengths for emission features in their spectra in conjunction with maps of reddening versus distance for stars observed along the line of sight to the nebula \citep*{lu73,ac78,ga86,pr92}. The recent IPHAS $H\alpha$ survey of the Galactic plane has generated extensive visible/infrared photometry for many stars in the fields of PNe, thereby permitting extension of the technique to a greater selection of PNe \citep{sa09}, as done by \citet{gi11}.

As a result of many such studies, distance estimates for PNe appear to fall into two categories: a ``short'' scale advocated by groups such as \citet{ca71}, \citet*{ka90}, and \citet{ph04}, and a ``long'' scale tied to the results of \citet{zh95} and \citet{gi11}. The ``short'' scale is supported by trigonometric parallaxes of PNe central stars \citep[e.g.][]{pa98,ha07,be09}, although the relative proximity to the Sun for such objects makes them less-than-ideal for such an assessment. A more definitive test could be made by membership of a PN in a distant open cluster, but, as noted above, the PN phase is relatively short-lived, more than 99\% of open clusters capable of housing their progenitors have dissolved by the time they are old enough to produce one, and most cases examined so far are line-of-sight projections rather than physical associations \citep{ma07}. 

A new case of a planetary nebula/open cluster association was recently discovered by \citet*{bo08} involving the planetary nebula Abell 8 (PN G167.0--00.9) with the cluster Bica 6. Abell 8 was discovered on Palomar Observatory Sky Survey images by \citet{ab55,ab66}, and the associated cluster by Bica \citep[see][]{bo08}; its present designation is from the 2009 update of the open cluster catalogue of \citet{di02}. Preliminary parameters for Bica 6 from 2MASS {\it JHK$_s$} data imply a cluster age of $1.0 \pm0.1$ Gyr \citep{bo08}, indicating a progenitor star of less than $\sim 3\,M_{\sun}$. It is a sparse cluster with an as-yet uncertain implied reddening and distance, a feature that might be resolved through a photometric and spectroscopic study at visible wavelengths. Presented here are new observations of that type revealing interesting characteristics of the cluster and its embedded planetary nebula, and strengthening the case for their association.

It is also noted here that the central star of Abell 8 appears to be double at far infrared wavelengths, a characteristic not seen in the visual region. That point requires further study, since the detection is at the brightness limits of the 2MASS survey.  

\begin{deluxetable}{@{\extracolsep{-2.5mm}}lcccclccc}
\tabletypesize{\tiny}
\tablewidth{0pt}
\tablecaption{Log of DAO photometric observations \label{tbl-1}}
\tablehead{
\colhead{Date (UT)} &\colhead{Filter} &\colhead{Exp. Time (s)} &\colhead{Air Mass (start)} & &\colhead{Date (UT)} &\colhead{Filter} &\colhead{Exp. Time (s)} &\colhead{Air Mass (start)}}  
\startdata
& &Bica 6 & & & & &NGC 2419 \\
\\
24 January 2009 &{\it U} &$3\times90$ &1.014 & &24 January 2009 &{\it U} &$2\times90$ &1.198 \\
                &     &$3\times360$ &1.017 & & & &$2\times500$ &1.176 \\
                &{\it B} &$3\times40$ &1.021 & & &{\it B} &$3\times60$ &1.250 \\
                &     &$3\times180$ &1.026 & & & &$3\times60$ &1.045 \\
                &{\it V} &$4\times20$ &1.032 & & & &$2\times240$ &1.220 \\
                &     &$3\times90$ &1.037 & & & &$1\times240$ &1.035 \\ 
26 January 2009 &{\it U} &$3\times90$ &1.019 & & &{\it V} &$3\times30$  &1.297 \\
                &     &$3\times360$ &1.015 & & & &$3\times30$  &1.063 \\
                &{\it B} &$3\times40$ &1.029 & & & &$2\times180$ &1.281 \\
                &     &$3\times180$ &1.024 & & & &$3\times180$ &1.052 \\
                &{\it V} &$3\times20$ &1.035 & &26 January 2009 &{\it U} &$2\times90$ &1.016 \\
                &     &$2\times90$ &1.032 & & &  &$3\times360$ &1.014 \\
& & & & & & &$2\times500$ &1.319 \\
& &PG 0220+132 & & & & &$2\times500$ &1.319 \\
& & & & & &{\it B} &$1\times40$ &1.022 \\
24 January 2009 &{\it B} &$3\times240$ &1.242 & & & &$1\times180$ &1.018 \\
                &{\it V} &$3\times180$ &1.232 & & & &$1\times240$ &1.411 \\
& & & & & &{\it V} &$1\times20$ &1.028 \\
& &SA 101 & & & & &$2\times90$ &1.024 \\
& & & & & & &$3\times180$ &1.460 \\
24 January 2009 &{\it U} &$1\times90$ &1.517 \\
                &     &$1\times500$ &1.518 & & & &GD71 \\
26 January 2009 &{\it U} &$1\times90$ &1.435 \\
                &{\it U} &$1\times360$ &1.432 & &26 January 2009 &{\it U} &$2\times90$ &1.212 \\
                &{\it B} &$2\times40$ &1.542 & & &{\it U} &$2\times360$ &1.196 \\
                &{\it V} &$2\times20$ &1.517 & & &{\it B} &$1\times180$ &1.225 \\
                &{\it V} &$2\times90$ &1.518 & & &{\it V} &$2\times20$ &1.246 \\
\\
& &PG0918+029 & & & & &G3-33 \\
\\
26 January 2009 &{\it U} &$1\times90$ &1.435 & &26 January 2009 &{\it B} &$3\times180$ &1.250 \\
                &{\it U} &$1\times360$ &1.432 & & &{\it V} &$3\times90$ &1.263 \\
                &{\it B} &$3\times180$ &1.250 \\
                &{\it V} &$1\times20$ &1.440 \\
                &{\it V} &$2\times90$ &1.445 \\
\enddata
\end{deluxetable}

\section{{\rm \footnotesize OBSERVATIONS AND DATA REDUCTIONS}} 
\subsection{{\rm \footnotesize PHOTOMETRIC AND SPECTROSCOPIC OBSERVATIONS}} 
Low dispersion CCD spectra (60 \AA~mm$^{-1}$) for bright members of Bica 6 were obtained in January 2009 and September 2010 using the Cassegrain spectrograph on the 1.85m Plaskett telescope of the Dominion Astrophysical Observatory. The same telescope was also used in direct imaging mode to obtain broad band, all-sky, CCD, {\it UBV} photometry for the cluster in January 2009, as well as CCD photometry on the Gunn system in February 2010, the latter to provide independent estimates for the magnitudes and colors of the planetary nebula's central star. Cousins system CCD {\it BV(RI)}$_C$ photometry was obtained in March 2010 for cluster stars brighter than $V = 18$ using the 0.35m robotic telescope at Sonoita Research Observatory. The latter observations are tied to numerous standards \citep{la83,la92} observed over a large range of air masses, and have an external zero-point uncertainty of about 0.02 magnitude. A program of regular observation of the cluster in the Johnson {\it V}-band was also made at the Abbey Ridge Observatory \citep{ma08} in order to search for variability among bright cluster members, but without positive detections to date.
\begin{figure}[!t]
  \epsscale{0.80}
  \plotone{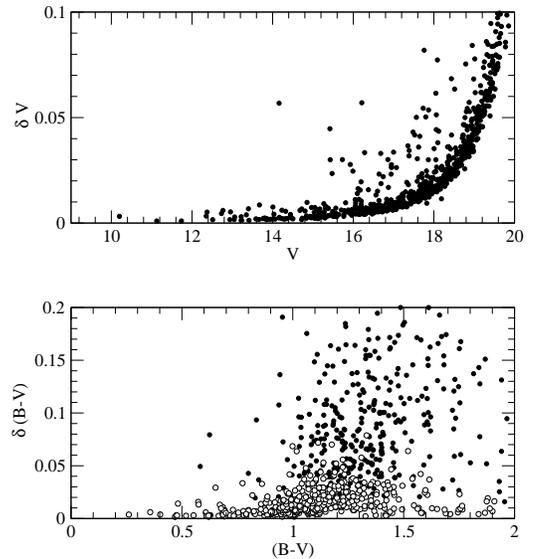}
  \caption{\small{Photometric uncertainties plotted as functions of magnitude and color. Open circles denote stars detected on at least four {\it B--}band frames.}}
  \label{fig1}
\end{figure}

\begin{figure}[!t]
  \epsscale{0.80}
  \plotone{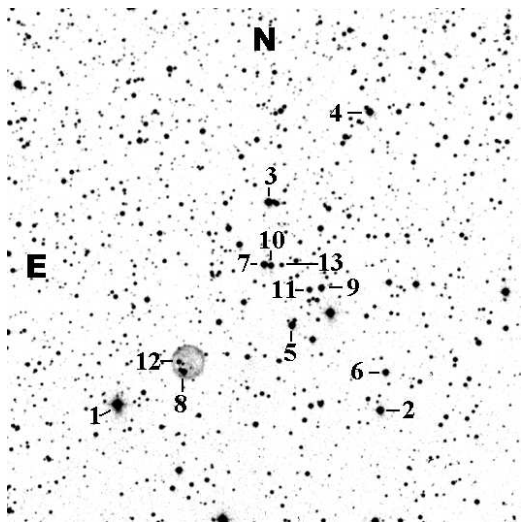}
  \caption{\small{A finder chart for the $15\arcmin$ diameter field of Bica 6 and Abell 8 centered on J2000 = 05:06:27.01, +39:11:01.7, derived from the POSS red image of the region, indicating stars observed spectroscopically in order of decreasing brightness.}}
  \label{fig2}
\end{figure}

\subsection{{\rm \footnotesize PHOTOMETRIC REDUCTIONS}} 
The optical {\it UBV} observations of Bica 6 were obtained on the nights of January 23/24 and January 25/26, 2009, PST with the 1.85m Plaskett telescope and E2V-1 CCD detector, both nights being considered to be of photometric quality. The CCD chip ($1024\times2304$ pixels binned $2\times2$) is too large to accommodate uniform illumination at the telescope focus, so the frames were trimmed to a useable size of $1020\times1800$ pixels before processing. The chip's image scale of 0.59 arcsec pixel$^{-1}$ and frame size after trimming resulted in a surveyed field measuring $10\arcmin\times17\arcmin.7$. 

A selection of \citet{la92} standard stars bracketing the air masses of the program fields and a standard field in NGC 2419 identified by Stetson (CADC local projects) were used for the transformations. Since Stetson's observations consist of {\it BVRI} only, photoelectric photometry from \citet{rh75} was used to obtain standard {\it U} magnitudes for NGC 2419. 

The images were bias-subtracted and flat-fielded using standard routines within IRAF. Absolute photometry was performed on all frames according to the usual procedure of point-spread function fitting using DAOPHOT and ALLSTAR, followed by aperture corrections and transformations to the standard system using routines provided by Stetson (private communication). Because of concerns about the photometric stability of the two nights of observation, transformation equations initially were obtained by allowing the software to compute the solution using individually-determined zero-points for the images. Good frames were selected on the basis of the zero-points and the FWHM measurements of the stars, as well as visual inspection of the frames themselves. Since the quality of the nights proved to be somewhat uneven, a selection of local standards were used as well; the Sonoita photometry was used for those stars. A log of the DAO observations that were used in the analysis is given in Table~\ref{tbl-1}. Note that, as a result of the above frame selection process, the program field air masses were not always bracketed by the standard ones.

The transformation equations used were of the form: 
\begin{eqnarray}
v & = & V + a_0 + a_1X + a_2(B-V) \nonumber \\
b & = & B + b_0 + b_1X + b_2(B-V) \nonumber \\
u & = & U + c_0 + c_1X + c_2(U-B) \nonumber \\
\nonumber \end{eqnarray}
where {\it b}, {\it v}, and {\it u} are the instrumental magnitudes, {\it B}, {\it V}, and {\it U} are the transformed magnitudes, {\it X} is the air mass, and the colors are the standard values. The transformation coefficients were obtained iteratively using least squares, with observations weighted according to the standard and observed photometric uncertainties and the residuals (observed minus actual). Typical weights ranged from 1.0 for residuals less than 0.022 to 0.10 for residuals of 0.042. Since the quality of the two nights differed, transformation coefficients for each night were obtained separately, and kept for the rest of the calibration (a useful feature of the Stetson routines); the photometry from both nights was combined in the last step (weighted according to the quality of the data). The coefficients obtained are in Table~\ref{tbl-2}. The average uncertainty (s.e.) per observation ranged from $\pm0.006$ for {\it V} to $\pm0.029$ for {\it U}.

\begin{deluxetable}{@{\extracolsep{-2.5mm}}lccclcc}
\tabletypesize{\scriptsize}
\tablewidth{0pt}
\tablecaption{Transformation coefficients for DAO observations \label{tbl-2}}
\tablehead{
\colhead{Date (UT)} &\colhead{Coefficient} &\colhead{Value} & &\colhead{Date (UT)} &\colhead{Coefficient} &\colhead{Value} } 
\startdata
24 January 2009 &a$_0$ &$1.815 \pm0.002$ & &26 January 2009 &a$_0$ &$1.750 \pm0.001$ \\
                &a$_1$ &$0.277 \pm0.006$ & & &a$_1$ &$0.183 \pm0.003$ \\
                &a$_2$ &$0.056 \pm0.003$ & & &a$_2$ &$0.068 \pm0.001$ \\
                &    &                    \\
                &b$_0$ &$2.027 \pm0.005$ & & &b$_0$ &$2.053 \pm0.004$ \\
                &b$_1$ &$0.323 \pm0.015$ & & &b$_1$ &$0.574 \pm0.007$ \\
                &b$_2$ &$-0.149 \pm0.006$ & & &b$_2$ &$-0.162 \pm0.003$ \\
                &    &                    \\
                &c$_0$ &$4.653 \pm0.006$ & & &c$_0$ &$4.630 \pm0.004$ \\
                &c$_1$ &$0.789 \pm0.027$ & & &c$_1$ &$0.650 \pm0.025$ \\
                &c$_2$ &$-0.079 \pm0.011$ & & &c$_2$ &$-0.075 \pm0.006$ \\
\enddata
\end{deluxetable}

Typical PSF-fitting uncertainties as determined by DAOPHOT and ALLSTAR ranged from $\delta u$, $\delta v$, $\delta b \simeq \pm0.001, \pm0.001, \pm0.002$ for instrumental magnitudes of $u$, $b$, $v$ = 12, to $\delta u$, $\delta b$, $\delta v \simeq \pm0.013, \pm0.024, \pm0.027$ for instrumental magnitudes of 16, for combined $U$, $B$, $V$ images of exposure time 360 s, 180 s, and 90 s, respectively. Uncertainties for the adopted aperture corrections varied between $\pm0.0002$ to $\pm0.0147$ depending on the aperture, with a median value of about $\pm0.002$. Uncertainties in the final, transformed photometry are presented in Fig.~\ref{fig1}.

\subsection{{\rm \footnotesize SPECTROSCOPIC REDUCTIONS AND SPECTRAL CLASSIFICATIONS}} 
\begin{figure}[!t]
  \epsscale{0.80}
  \plotone{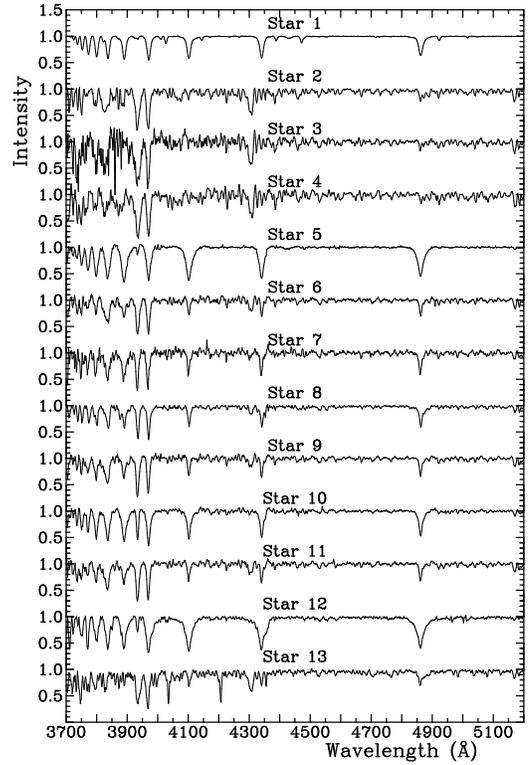}
  \caption{\small{A montage of normalized spectra for Bica 6 stars identified in Fig.~\ref{fig2}.}}
  \label{fig3}
\end{figure}

The spectra were used to classify cluster stars on the MK system using standard techniques involving line strengths and line ratios in the blue-green spectral region \citep[see][]{gc09}, as well as to derive radial velocities for the stars. A finder chart for the spectroscopically observed objects is presented in Fig.~\ref{fig2}, the continuum-flattened spectra are plotted in Fig.~\ref{fig3}, and Table \ref{tbl-3} summarizes information for the stars, as well as for the central star of Abell 8. The stars are identified by their J2000 co-ordinates in the 2MASS catalogue \citep{cu03}, converted to degrees, and the numbering of Fig.~\ref{fig2}. The data include estimates of {\it V}, {\it B--V}, and {\it U--B}, with uncertainties of less than $\pm0.01$ magnitude for {\it V} and {\it B--V} (except for the central star), spectral types assigned with reference to the MK spectral atlas as well as \citet{gc09}, reddening and distance modulus information inferred from the spectral classifications, and radial velocities derived from parabolic fits to the cores of reasonably strong, unblended lines in the spectra (54 lines per star, on average).

The precision of stellar radial velocities depends upon the spectral dispersion \citep[e.g.][]{sc83}, and for our wavelength-calibrated 60 \AA~mm$^{-1}$ spectra is typically $\pm 10$ km s$^{-1}$ or better (p.e.). The accuracy of the radial velocities obtained in this fashion is generally more secure. For example, a spectral scan of a 103 \AA~mm$^{-1}$ spectrogram for the Hyades spectroscopic binary HD 30869 by \citet*{te86} measured in such fashion yielded a velocity within 2 km s$^{-1}$ of the actual value from only 5 unblended lines. Tests using a variety of CCD spectra obtained with the Plaskett telescope's Cassegrain spectrograph measured in similar fashion indicate no need for a zero-point adjustment.

Star 1 is HDE 277593, classified as B8e in the {\it Henry Draper Extension}. Our spectrogram does not extend to $H\alpha$, so we cannot confirm the emission-line nature of the star; higher Balmer series members are all in absorption. Moreover, the spectral lines display strong rotational broadening, and also indicate a much earlier spectral type: B1~Vnn. Such a hot star appears anomalous, lying in the midst of a cluster of F-type and G-type giants. Yet its radial velocity is comparable to the velocities of other cluster members, which are about 60 km s$^{-1}$ more positive than expected for standard Galactic rotation. The mean cluster velocity for the eleven stars of similar radial velocity is 57.1 $\pm1.2$ km s$^{-1}$, where the cited uncertainty represents the standard error of the mean. The two deviant objects are star 7, conceivably a spectroscopic binary given its similar reddening and distance modulus to other cluster stars, and star 13, which has a velocity consistent with Galactic rotation for a foreground star.

Table \ref{tbl-3} includes information on the reddening and distance moduli of the stars, most of which are more luminous than zero-age main sequence (ZAMS) objects, where intrinsic colors were adopted from \citet{fi70}, \citet{kr78}, and, for main-sequence stars, stellar atmosphere predictions for the colors of zero-age zero-rotation stars \citep{cs85} with a zero-point adjustment to match A0 stars \citep[see][]{tu92}, and absolute visual magnitudes from the compilation of \citet{gc09} in conjunction with an extension of the calibration of \citet{tu80}. A ratio of total-to-selective extinction of $R = A_V/E_{B-V} = 3.0 \pm0.1$ and a reddening relation of slope $E_{U-B}/E_{B-V}=0.75$ for early-type stars were found previously for fields lying at the Galactic longitude of Bica 6 \citep{tu76b}, and were adopted here. The variation of apparent distance modulus, {\it V--M}$_V$, with color excess, $E_{B-V}$, for stars in Table \ref{tbl-1} is consistent with such a result (see \S3).

\begin{deluxetable}{@{\extracolsep{-2.5mm}}cccccclccccccc}
\tabletypesize{\scriptsize}
\tablewidth{0pt}
\tablecaption{Spectroscopic and Photometric Data for Bright Bica 6 Stars \label{tbl-3}}
\tablehead{
\colhead{Star} &\colhead{RA(2000)} &\colhead{Dec(2000)} &\colhead{{\it V}} &\colhead{{\it B--V}}  &\colhead{{\it U--B}} &\colhead{Sp.T.} &\colhead{{\it B--V}$_0$} &\colhead{$E_{B-V}$} &\colhead{{\it V--M}$_V$} &\colhead{{\it V}$_0${\it --M}$_V$} &\colhead{$V_R$} &\colhead{$\Delta V_R$} &\colhead{Lines} \\
& & & & & & & &(B0) & & &km s$^{-1}$ &km s$^{-1}$ } 
\startdata
1\tablenotemark{a} &76.70364 &39.11653 &10.20 &0.26 &--0.52 &B1 Vnn &--0.27 &0.53 &12.33 &10.73 &+61 &$\pm6$ &24 \\
2 &76.53968 &39.11315 &12.35 &1.43 &+1.07 &G6 II-III &+0.92 &0.56 &12.85 &11.18 &+55 &$\pm10$ &60 \\
3 &76.60863 &39.21426 &12.51 &1.37 &+1.24 &G8 III &+0.95 &0.47 &12.21 &10.81 &+53 &$\pm10$ &56 \\
4 &76.54565 &39.25783 &12.69 &1.37 &+1.30 &G7 III &+0.94 &0.46 &12.39 &11.02 &+56 &$\pm6$ &50 \\
5 &76.59442 &39.15476 &12.73 &0.36 &+0.12 &B9.5 IV &--0.03 &0.39 &12.43 &11.24 &+61 &$\pm7$ &56 \\
6 &76.53616 &39.13146 &12.94 &0.76 &+0.26 &F7 III &+0.50 &0.27 &12.34 &11.52 &+55 &$\pm7$ &63 \\
7 &76.61181 &39.18389 &12.95 &0.78 &+0.33 &F2 III &+0.36 &0.44 &12.35 &11.04 &(+23) &$\pm11$ &62 \\
8 &76.66211 &39.13222 &13.05 &0.71 &+0.23 &F3 III &+0.38 &0.36 &12.45 &11.38 &+62 &$\pm11$ &45 \\
9 &76.57605 &39.17258 &13.06 &0.78 &+0.31 &F5 III &+0.41 &0.39 &12.36 &11.19 &+51 &$\pm8$ &68 \\
10 &76.60746 &39.18347 &13.24 &0.62 &+0.42 &F0 IV &+0.31 &0.33 &11.54 &10.56 &+59 &$\pm6$ &26 \\
11 &76.58347 &39.17167 &13.88 &0.69 &+0.41 &F6 IV &+0.46 &0.25 &11.68 &10.93 &+54 &$\pm6$ &73 \\
12 &76.66512 &39.13706 &14.72 &0.67 &+0.43 &A3 V &+0.09 &0.59 &12.72 &10.94 &+60 &$\pm5$ &45 \\
13 &76.60073 &39.18366 &15.59 &0.99 &+0.34 &G0 V: &+0.59 &0.43 &11.99 &10.71 &(--2) &$\pm6$ &96 \\
CS &76.65983 &39.135833 &19.93 &0.15 &\nodata &\nodata &(--0.38) &(0.49) &\nodata &(11.02) &\nodata &\nodata &\nodata \\
\\					
Average & & & & & & & &0.42 & &11.02 &+57 \\
s.e. & & & & & & & &$\pm0.03$ & &$\pm0.08$ &$\pm1$ \\
\enddata
\tablenotetext{a}{HDE 277593, B8e.}
\end{deluxetable}

\subsection{{\rm \footnotesize PHOTOMETRIC STANDARDIZATION}} 
As noted above, the direct images of Bica 6 from the DAO were calibrated using comparable images of standard stars observed at air masses similar to those for Bica 6. For consistency, minor adjustments were made to the {\it BV} magnitudes to improve the match to the all-sky CCD {\it BV(RI)}$_C$ photometry from the Sonoita images. A similar procedure was used for the initial calibration of the {\it U-}band images, namely by reference to {\it U} magnitudes for standard stars. However, most current CCD-telescope-filter combinations produce very poor matches to the wavelength sensitivity of the Johnson {\it U} filter \citep[see][]{tu11}, and in such cases the calibration of {\it U} magnitudes and {\it U--B} colors can be non-linear and possibly multi-valued \citep{mv77,cc01}. In fact, the initial {\it UBV} color-color diagram constructed for cluster stars displayed characteristics indicating systematic effects in the {\it U} data.

\begin{figure}[!t]
  \epsscale{0.80}
  \plotone{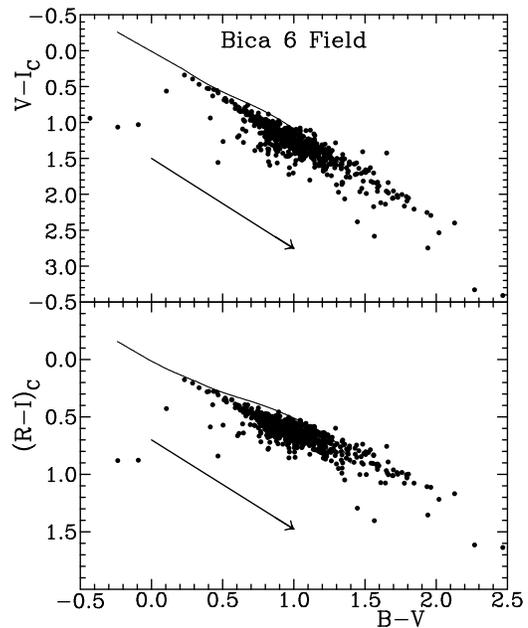}
  \caption{\small{{\it V--I}$_C$ and {\it (R--I)}$_C$ colors as a function of {\it B--V} for stars in the Bica 6 field relative to the intrinsic relations for dwarfs (curved relations). Arrowed lines depict the effects of interstellar reddening for $E_{B-V} = 1.00$.}}
  \label{fig4}
\end{figure}

Removal of such effects from the final data set was done as follows. The {\it BV(RI)}$_C$ colors for Bica 6 stars plotted in Fig.~\ref{fig4} indicate that the only unreddened objects in the field are late-type stars and one or two likely early A-type stars. The majority of stars in the sample are reddened by about $E_{B-V}\simeq0.5$. For spectroscopically-observed stars, one can relate the spectral type of the star to tabulated intrinsic colors \citep[e.g.,][]{fi70} to calculate the correction, $\delta${\it (U--B)}, to the initially observed {\it U--B} color, required to make the computed color excesses $E_{U-B}$ and $E_{B-V}$ fit the accepted reddening relation for the field (see above), adjusted in the case of the cooler stars for the color dependence of color excesses and reddening slope \citep[e.g.,][]{fe63}. For a few stars it was possible to identify them from Fig.~\ref{fig4} as unreddened main-sequence stars from their faintness and {\it BV(RI)}$_C$ colors, thereby indicating the correction to {\it U--B} needed to make their colors fit the intrinsic color-color relation for dwarfs. Finally, the field reddening for spectroscopically-observed stars varies relatively smoothly across the field, so some of the fainter stars lying within roughly $30 \arcsec$ of them and indicated to be reddened from their {\it BV(RI)}$_C$ colors were assumed to share the same reddening, as an approximation, thereby allowing additional corrections to {\it U--B} to be calculated, as above.

\begin{figure}[!t]
  \epsscale{0.80}
  \plotone{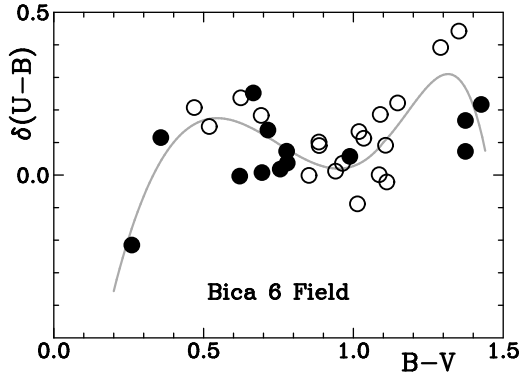}
  \caption{\small{The correction for {\it U--B} colors as a function of {\it B--V} for stars observed spectroscopically (filled circles) and stars that are either unreddened or reddened as their nearby spectroscopic neighbors (open circles). The plotted gray relation was adopted to adjust the CCD colors to the Johnson system.}}
  \label{fig5}
\end{figure}

The dependence of the derived corrections for the stars is plotted as a function of observed {\it B--V} color in Fig.~\ref{fig5}, producing a reasonably well-defined dependence matched to a best-fitting polynomial. Normally such corrections are spectral type dependent \citep[see][]{mv77}, but most stars in the Bica 6 field are reddened by similar amounts, which are not large, so a simple color dependence appears to be appropriate. The adopted color dependence $\delta${\it (U--B)} shown in Fig.~\ref{fig5} was therefore applied to all observed stars, except very red objects with $B-V \ge 1.45$, for which a constant correction of +0.15 produced more reasonable results.

\begin{figure}[!t]
  \epsscale{0.80}
  \plotone{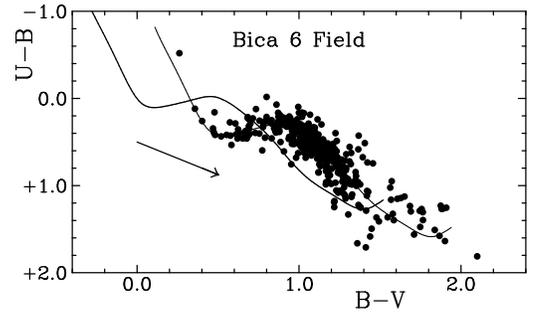}
  \caption{\small{Color-color diagram, {\it U--B} versus {\it B--V}, for stars in Bica 6, from the final adjusted photometry. The solid curve represents the intrinsic relation for main-sequence stars, the narrow curve the same relation reddened by $E_{B-V}=0.42$. An arrowed line shows the effects of interstellar reddening on the colors for $E_{B-V}=0.5$.}}
  \label{fig6}
\end{figure}

The effectiveness of such a procedure can be judged from Fig.~\ref{fig6}, which plots the {\it UBV} color-color diagram for the 391 stars in the final adjusted data set that have full {\it UBV} coverage. The colors for cluster stars are a good match to the intrinsic relation reddened by $E_{B-V}=0.42$, which represents the mean color excess for the field. According to the results of Fig.~\ref{fig5}, the uncertainties in the adjusted {\it U--B} colors may be as large as $\pm0.10$, although for most stars they are probably not much larger than $\pm0.05$. Within such limits, the color-color data of Fig.~\ref{fig6} are well matched to the reddened intrinsic relation over the entire range of likely spectral types for the stars, from B-type stars to M-type stars.

The final corrected {\it UBV} observations for Bica 6 stars, as well as {\it (RI)}$_C$ data, are on deposit at the WEBDA Open Cluster Database, http://www.univie.ac.at/webda/, and identify the 2000 co-ordinates for each star along with the observed values of {\it V}, {\it B--V}, {\it U--B}, {\it V--R}$_C$, {\it V--I}$_C$, and {\it (R--I)}$_C$, and their uncertainties. The complete table of CCD data for almost 800 observed stars is too large to publish here.

\section{{\rm \footnotesize PARAMETERS OF BICA 6}} 
\begin{figure}[!t]
  \epsscale{0.80}
  \plotone{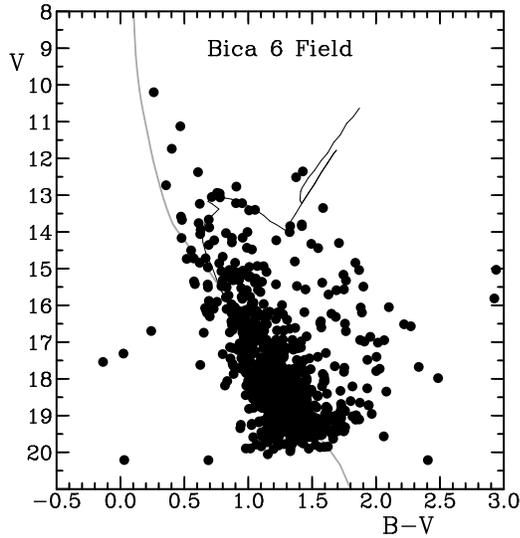}
  \caption{\small{Color-magnitude diagram for Bica 6 from the full data set. The gray curve is the ZAMS for $E_{B-V}=0.42$ and {\it V--M}$_V=12.28$; the black curve is a theoretical isochrone for $\log t=9.0$ from \citet{me93} for the same parameters.}}
  \label{fig7}
\end{figure}

Fig.~\ref{fig7} is a color-magnitude diagram for the complete data set of observed Bica 6 stars, with no elimination of foreground or background stars. Included is the zero-age main sequence (ZAMS) relation from \citet{tu76a,tu79} reddened by $E_{B-V}=0.42$ for {\it V--M}$_V=12.28$, along with a theoretical isochrone for $\log t=9.0$ from \citet*{me93} matched to the same parameters. The parameters are in very close agreement with the results obtained by \citet{bo08} from 2MASS {\it JHK}$_s$ data. A variety of more recently published isochrones could have been plotted for comparison, with nearly identical results. Those of \citet{me93} were employed because they appear to provide slightly better fits to the observational data for stars in open clusters than is the case with some others \citep[see][]{tu11}.

\begin{figure}[!t]
  \epsscale{0.80}
  \plotone{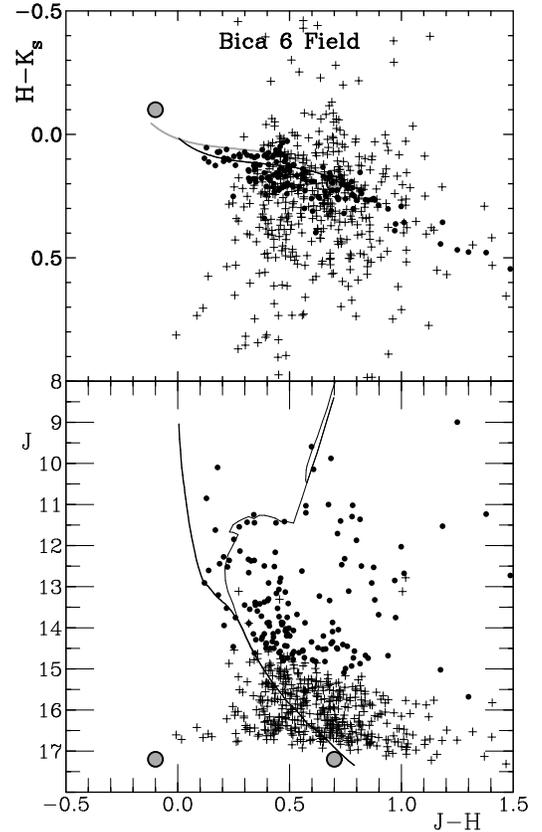}
  \caption{\small{2MASS color-color diagram, {\it H--K}$_{\rm s}$ versus {\it J--H} (upper), and color-magnitude diagram, {\it J} versus {\it J--H} (lower), for stars within $6\arcmin$ of the center of Bica 6, from observations by \citet{cu03}. In the upper plot, the intrinsic color-color relation for main sequence stars is plotted as a gray curve, while a solid black curve is the same relation reddened by $E_{J-H}=0.124$ ($E_{B-V}=0.42$). In the lower plot, the ZAMS is plotted as a black curve for the same reddening and {\it J--M}$_J=11.32$ ({\it V--M}$_V=12.28$), while the gray curve is an isochrone for $\log t=9.0$. Filled circles represent data with uncertainties smaller than $\pm0.06$, plus signs represent data with larger uncertainties, and gray circled points are estimates for the two stars near the center of Abell 8.}}
\label{fig8}
\end{figure}

An independent solution for Bica 6 was attempted with the 2MASS observations \citep{cu03} using the intrinsic {\it JHK}$_s$ relations of \citet{tu11}. {\it JHK}$_s$ data for stars lying within $6\arcmin$ of the center of Bica 6 cited by \citet{bo08} are plotted in Fig.~\ref{fig8} along with the intrinsic relations of \citet{tu11}. The data were separated according to the cited uncertainties in the magnitudes into a high quality group, $\sigma \le \pm0.06$, and a low quality group with larger uncertainties. It was difficult to establish the reddening uniquely, either because of large scatter in the data or because of differential reddening in the field, although there is a reasonably good match of the data to $E_{J-H}=0.124$, corresponding to the average reddening of $E_{B-V}=0.42$ obtained for the {\it UBV} data. For that reddening the apparent distance modulus is {\it J--M}$_J=11.32$ from ZAMS fitting, coincident with the distance modulus of {\it V--M}$_V=12.28$ used with the {\it BV} data in Fig.~\ref{fig7}. There is also a close match of a $\log t=9.0$ isochrone to the {\it JHK}$_s$ data, as noted by \citet{bo08}, although the model isochrone used in Fig.~\ref{fig8} is an independent relation adapted from \citet{me93} \citep[see][]{tu11}.

\begin{figure}[!t]
  \epsscale{0.80}
  \plotone{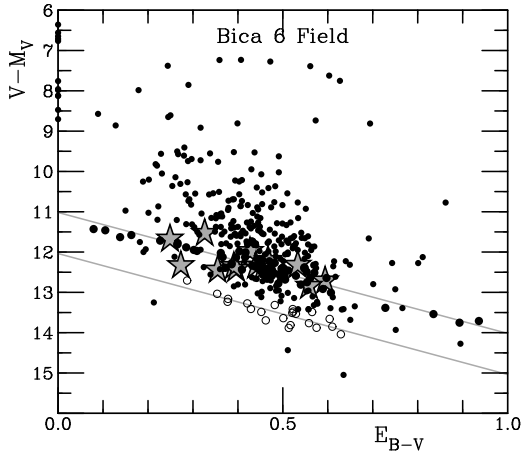}
  \caption{\small{Variable-extinction diagram for stars in the Bica 6 field (filled circles), including likely ZAMS members (large filled circles), likely ZAMS stars for the background group (open circles), and spectroscopic distance moduli for stars in Table~\ref{tbl-1} (gray star symbols). The two gray lines are $R=3.0$ relations for the Bica 6 cluster (upper) and the background group (lower).}}
\label{fig9}
\end{figure}

The existence of differential reddening across the field of Bica 6 is evident from the reddenings for spectroscopically-observed stars (Table~\ref{tbl-3}), as well as from sizeable scatter in the {\it UBV} and 2MASS colors for cluster stars (Figs.~\ref{fig6}~\&~\ref{fig8}) and visual inspection of images of the field. There is a noticeable increase in the visual extinction on the southwest side of the field, according to an apparent decrease in star densities there, and that is confirmed by the derived color excesses of stars in the region. As a consequence, the final parameters for Bica 6 were established from a variable-extinction analysis of the {\it UBV} observations for cluster stars \citep[see][]{tu76a,tu76b}. Stars with full {\it UBV} coverage were therefore individually dereddened to the intrinsic relation for main-sequence stars in Fig.~\ref{fig6}, color excesses were adjusted to equivalent values for a B0 star using the relationship of \citet{fe63}, and apparent distance moduli were calculated using ZAMS values of {\it M}$_V$ from \citet{tu76a,tu79}. The results are shown in Fig.~\ref{fig9} along with the apparent distance moduli for spectroscopically observed stars in Table~\ref{tbl-3}.

An alternate method to dereddening the data is possible using {\it BVRI}$_C$ or other infrared colors, but such an approach in this instance would result in greater uncertainties in the derived color excesses. Most of the stars in the field are of apparently late spectral type, for which {\it BVRI}$_C$ reddening lines run nearly parallel to the intrinsic color relations for dwarfs \citep[see][and Fig.~\ref{fig4}]{te11}. Attempts to infer intrinsic colors for stars in the observational sample would therefore produce color excesses $E_{B-V}$ of larger uncertainty than those generated from analysis of the {\it UBV} observations.

An initial concern with such an approach was the possibility that large scatter in the {\it U--B} colors might produce biased results. The effects of random photometric errors on variable-extinction analyses of O-type, B-type, and A-type stars were investigated previously by \citet{tu76a}, and it is a simple matter to extend the results to the GK dwarfs dominating the field of Bica 6. As a consequence, we expect the determination of $R=A_V/E_{B-V}$ for cluster stars to be biased towards a value of $R=1.9$, with a total spread in color excesses $E_{B-V}$ approximating $\sim0.2$. The data of Fig.~\ref{fig9} are at complete variance with such expectations. There is no apparent bias towards a value of $R\simeq2$, and the spread of color excesses is more than twice as large as expected. There are also many unreddened stars in the field. Given the generally consistent trends of the data in both Fig.~\ref{fig6} and Fig.~\ref{fig9} to what is typical of the scatter in cluster color-color and variable-extinction diagrams for good quality data \citep[see][]{tu76a,tu76b}, it appears that the uncertainties in the derived {\it U--B} colors amount to $\pm0.10$ only in extreme cases. The true scatter for most cluster stars appears to be much smaller, as noted earlier.

The data of Fig.~\ref{fig9} fall into five distinct groups: (i) unreddened foreground stars with apparent distance moduli of $V_0${\it --M}$_V\le8.7$, indicating that the dust clouds creating much of the extinction in the Bica 6 field lie at distances of $d\simeq550$ pc, (ii) reddened foreground stars more distant than 550 pc with color excesses ranging from 0.2 to 0.9 and apparent distance moduli of 7--12, (iii) three apparently reddened background objects (or stars with anomalous data) with true distance moduli of $V_0${\it --M}$_V\ge12.5$, (iv) a group of 64 stars with $V_0${\it --M}$_V\simeq11$ and color excesses ranging from 0.08 to 0.94, most of which are likely ZAMS members of Bica 6, and (v) a group of 22 stars with $V_0${\it --M}$_V\simeq12$ and color excesses ranging from 0.29 to 0.63 that appear to be ZAMS members of a group lying background to Bica 6. All of the stars in the last group (v) lie on the northeast side of the field, and may be associated with a sparse, perhaps dissolved, cluster of faint stars that appears to lie in that region. The data for stars in group (v) were analyzed using least squares and non-parametric techniques, yielding a value of $R=A_V/E_{B-V}=3.05\pm0.31$ s.e. and $V_0${\it --M}$_V=12.01\pm0.15$ s.e., corresponding to a distance of $2.56\pm0.18$ (s.d.) kpc. 

The data for stars in the more populous group (iv) were also analyzed using least squares and non-parametric techniques, which yielded a value of $R=A_V/E_{B-V}=2.99\pm0.13$ s.e. and $V_0${\it --M}$_V=11.02\pm0.06$ s.e., corresponding to a distance of $1.60\pm0.11$ (s.d.) kpc. The values of {\it R} derived for both groups agree with prior expectations ($R=3.0$, \S2.3) for the characteristic properties of dust clouds lying in the direction of Bica 6 \citep{tu76b}, and the implied distance of $\sim550$ pc to the dust complex along the line of sight matches independent results by \citet{nk80} for the distribution of dust along adjacent Galactic sight lines. The latter results also indicate that there is no additional reddening in this direction for a further 2 kpc or so, presumably implying that the interarm region between the local Cygnus arm and the Perseus structure is relatively free of dust and extinction. That would explain why there is no obvious separation of groups (iv) and (v) in Fig.~\ref{fig9} in terms of color excess, $E_{B-V}$. The stars in both groups have been reddened by the dust associated with a complex lying well foreground to them.

It is worth pointing out that results such as those presented here appear typical of what is found in other Galactic star fields associated with open clusters \citep{tu94}. Galactic dust complexes appear to be spatially restricted around the Sun \citep{nk80}, producing interstellar reddening and extinction of differing amounts for all stars lying beyond a dust complex or between individual complexes. The notion of interstellar reddening increasing monotonically with distance from the Sun is incorrect. In most directions the reddening is produced in one or more dust clouds lying along the line of sight, with the reddening for individual stars reflecting both the number of dust clouds lying between the star and the Sun and the amount of extinction occurring in each cloud along the line of sight to the star. In the case of the Bica 6 field, there appears to be only one dust complex detected to the limits of the survey, and it is located $\sim550$ pc distant.

Spectroscopically observed stars confirm the results of the variable-extinction analysis. The derived mean intrinsic distance modulus for stars in Table~\ref{tbl-3} is $V_0${\it --M}$_V=11.02\pm0.08$ s.e., identical to what was found from ZAMS fitting, regardless of questions about the membership status of individual stars. The corresponding distance of $1.60\pm0.06$ kpc agrees with the value of $1.7\pm0.1$ kpc obtained by \citet{bo08} for Bica 6 using an independent main sequence calibration. The close agreement of {\it UBV}, {\it JHK}$_s$, and spectroscopic distance estimates for Bica 6 provides a solid foundation for the distance of the cluster. The reddening is also confirmed to be differential across the face of the cluster, but lies mainly in the range from 0.2 to 0.6 in $E_{B-V}$, averaging 0.42 in $E_{B-V}$.

\section{{\rm \footnotesize CLUSTER MEMBER ABELL 8}} 
Some of the observed parameters of Abell 8 (PN G167.0--00.9) are relatively poorly established \citep*{ka83,ca92}, but sufficient to permit estimates of reddening from the H$\beta$/radio flux method \citep{ma75,ca76,bg85} or the Balmer line method \citep{of06}. The color excess is $E_{B-V} = 0.55 \pm0.45$ when one combines the 1.4 GHz flux density measured by \citet{ck98} with the H$\beta$ flux estimated by \citet*{ka90} and \citet{ca92}, the generous uncertainty resulting from propagation of the large relative uncertainties in the input quantities. Estimates for the reddening from the Balmer decrement \citep*{ca71,al99,ph05,gi11} in conjunction with the conversion factor of \citet{ma07} imply values of $E_{B-V} = 0.51$, 0.44, 0.42, and 0.47, respectively, with expected uncertainties for the latter of order $\pm0.08$ \citep[e.g.,][]{of06}. The reddening estimates are summarized in Table~\ref{tbl-4} along with independent estimates for radial velocity and distance.

\begin{deluxetable}{lllll}
\tabletypesize{\scriptsize}
\tablewidth{0pt}
\tablecaption{Deduced Parameters for Abell 8 and Bica 6 \label{tbl-4}}
\tablehead{
\colhead{Parameter} &\colhead{Abell 8} &\colhead{Source} &\colhead{Bica 6} &\colhead{Source} }
\startdata
$V_R$ &$58.2 \pm6.5$ km s$^{-1}$ &\citet*{pp81} &$57.1 \pm1.2$ km s$^{-1}$ &This paper \\
\\
$E_{B-V}$ &0.51\tablenotemark{a} &\citet{ca71} &$0.42\pm0.20\tablenotemark{e}$ &This paper \\
\nodata &0.44\tablenotemark{a} &\citet{al99} &\nodata &\nodata \\
\nodata &0.42\tablenotemark{a} &\citet{ph05} &\nodata &\nodata \\
\nodata &0.47\tablenotemark{a} &\citet{gi11} &\nodata &\nodata \\
\nodata &$0.55 \pm0.45$\tablenotemark{b} &This paper &\nodata &\nodata \\
\nodata &$0.48 \pm0.05$\tablenotemark{c} &This paper &\nodata &\nodata \\
\nodata &$0.53 \pm0.05$\tablenotemark{d} &This paper &\nodata &\nodata \\
\\
Distance &2.05 kpc &\citet{ab66} &$1.60\pm0.11$ kpc &This paper \\
\nodata &1.78 kpc &\citet{ca71} &\nodata &\nodata \\
\nodata &2.66 kpc &\citet{ka83} &\nodata &\nodata \\
\nodata &2.60 kpc &\citet{ma84} &\nodata &\nodata \\
\nodata &1.67 kpc &\citet{ka90} &\nodata &\nodata \\ 
\nodata &1.86 kpc &\citet{ca92} &\nodata &\nodata \\
\nodata &3.23 kpc &\citet{zh95} &\nodata &\nodata \\ 
\nodata &1.78 kpc &\citet{ph04} &\nodata &\nodata \\
\nodata &5.50 kpc &\citet{gi11} &\nodata &\nodata \\
\enddata
\tablenotetext{a}{Balmer decrement method.}
\tablenotetext{b}{$H\beta$/radio flux method.}
\tablenotetext{c}{Space reddening.}
\tablenotetext{d}{Spectroscopic/photometric reddening.}
\tablenotetext{e}{Range for most cluster members.}
\end{deluxetable}

\begin{figure}[!t]
  \epsscale{0.80}
  \plotone{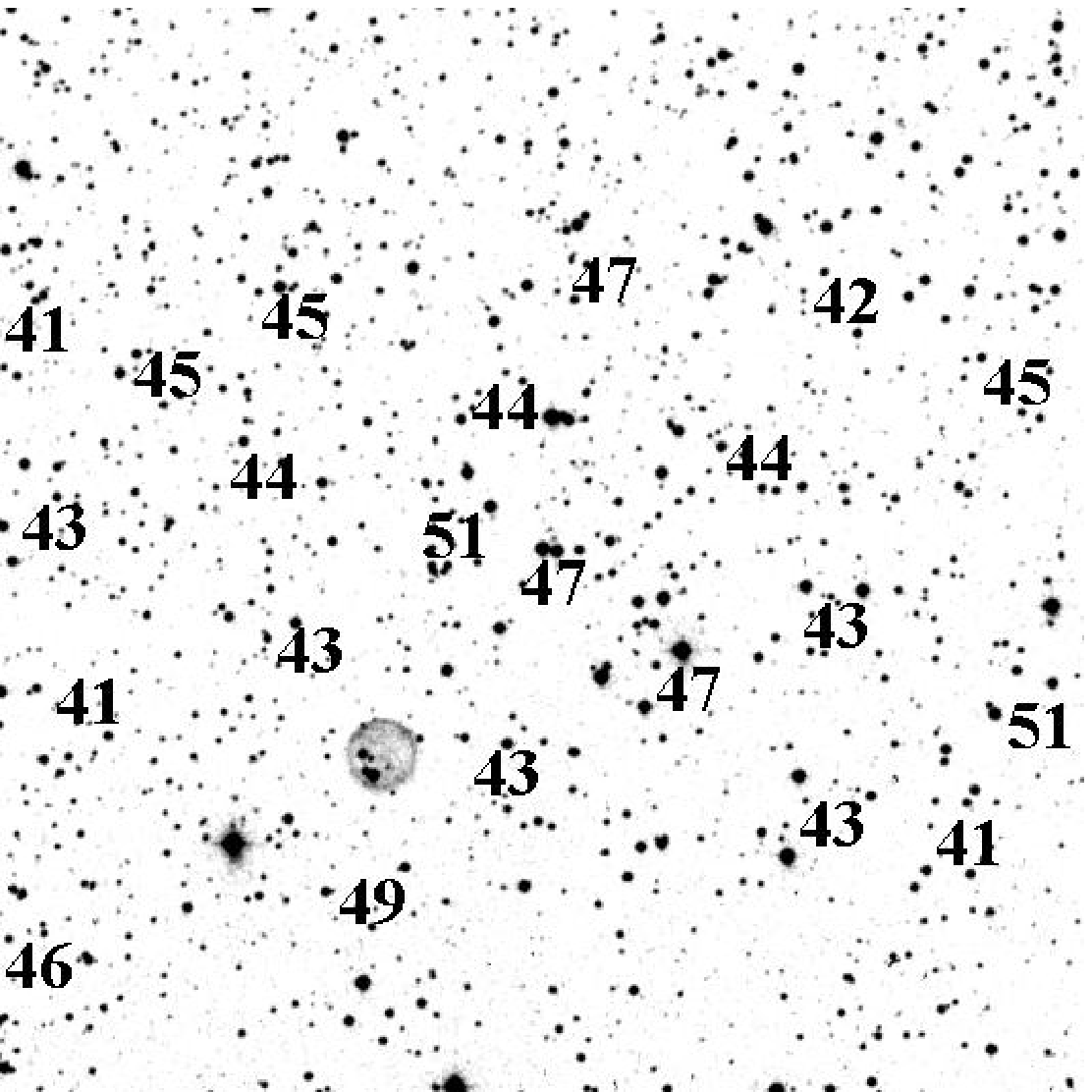}
  \plotone{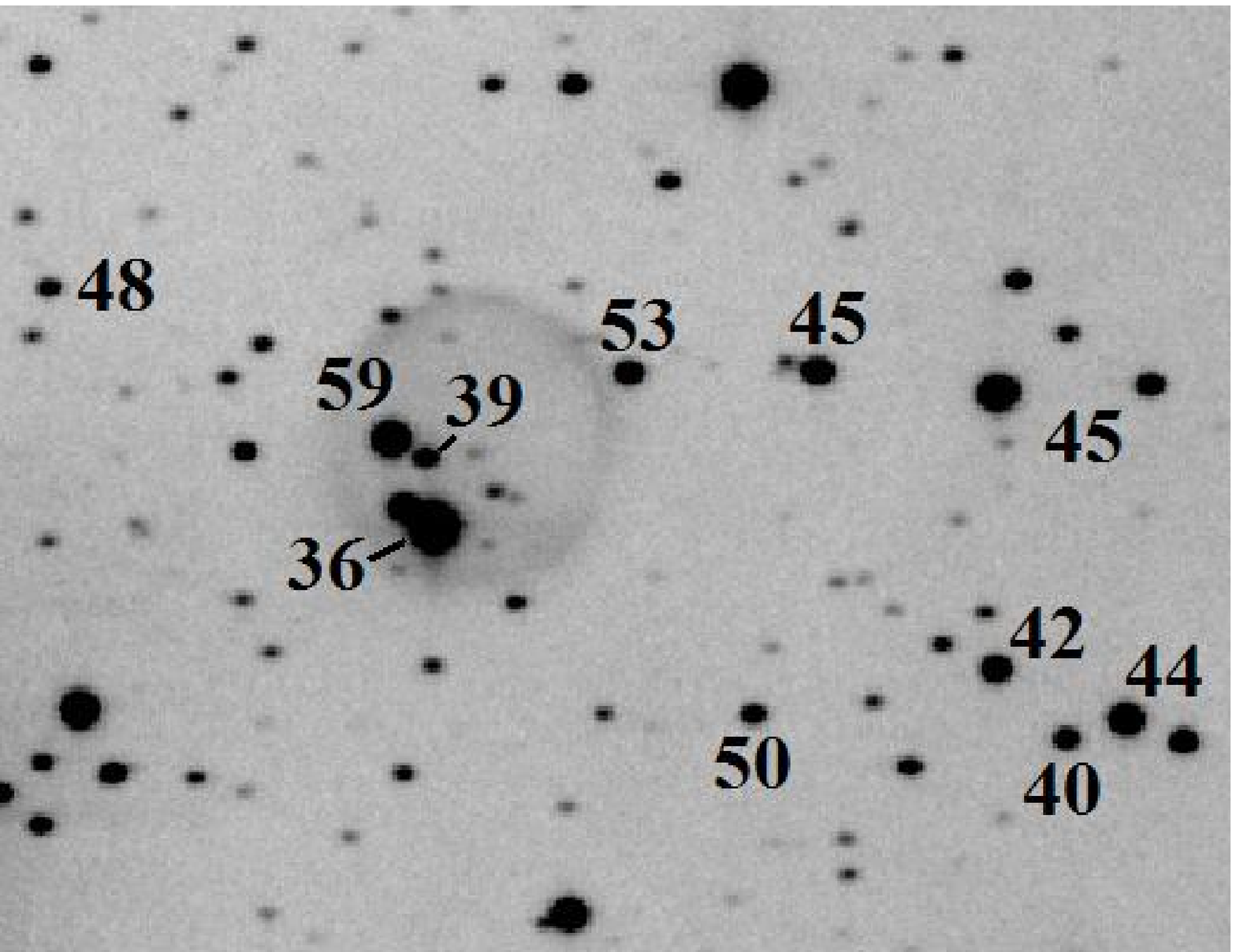}
  \caption{\small{Derived space reddenings (numerals in black) for Bica 6 stars for the field of Fig.~\ref{fig2} (upper) and in the vicinity of Abell 8 (lower), with reddenings cited in units of 0.01 magnitude without the decimal. The reddenings for Bica 6 stars represent group means for clumps of 6--8 stars in the same direction, those for Abell 8 are individual color excesses.}}
\label{fig10}
\end{figure}

An independent estimate is possible from the space reddening information obtained from the variable-extinction analysis presented here. The reddening of stars in Bica 6 and near Abell 8 (Fig.~\ref{fig10}) suggests the existence of internal reddening within the nebula itself. Stars near the planetary nebula have color excesses of $E_{B-V}= 0.37\pm0.01$ (2 in front), 0.59 (behind), and 0.53 (off west limb), implying an internal reddening of $E_{B-V}\simeq0.22$. Such a result appears reasonable given possible excess reddenings for PNe associated with dust produced during evolution on the asymptotic giant branch for their progenitor stars \citep[see][]{ma07}. The implied reddening of the central star is therefore $E_{B-V}= 0.48\pm0.05$, where the uncertainty reflects the full spread in internal reddening inferred for the planetary. A similar value of $E_{B-V}= 0.53\pm0.05$ is obtained by adopting an intrinsic {\it B--V} color of --0.38 from \citet{kf85} with the observed color from Table~\ref{tbl-5}. The various estimates of reddening are remarkably consistent, yielding a mean value of $E_{B-V}= 0.49\pm0.02$, adopted here.

The measured radial velocity of Abell 8 is $58.2 \pm6.5$ km s$^{-1}$ \citep*{sc83,du98}, according to measurements from low quality spectra by \citet{pp81}. That agrees with the radial velocity for Bica 6 members of $57.1 \pm1.2$ km s$^{-1}$ (Table~\ref{tbl-3}). The good agreement in the estimates of reddening and radial velocity for the planetary nebula with values obtained for stars in Bica 6 provides reasonably good evidence for their physical association, although the case from reddening alone is clearly weak, given that it results from dust clouds lying well foreground to Bica 6. 

Estimates for the distance to Abell 8 (PN G167.0--00.9) are typically linked to inferences about its reddening, but vary from ``short'' values of 1.67 kpc \citep{ka90} and 1.78 kpc \citep{ca71,ph04} to ``long'' values of 3.23 kpc \citep{zh95} and 5.5 kpc \citep{gi11} according to the methodology used, as indicated in Table~\ref{tbl-4}. The former are very close to the distance of $1.60 \pm0.11$ kpc obtained here for Bica 6, thereby favoring the ``short'' scale of planetary nebula distances.

\begin{deluxetable}{ccl}
\tabletypesize{\scriptsize}
\tablewidth{0pt}
\tablecaption{The Central Star of Abell 8 \label{tbl-5}}
\tablehead{
\colhead{{\it V}} &\colhead{{\it B--V}} &\colhead{Source} }
\startdata
19.89 &0.15 &This paper ({\it BV} photometry) \\
19.99 &0.10 &This paper (Gunn system photometry) \\
19.9 &0.2 &POSS; \citet{ab66,ka90} \\
\\
19.93 &0.15 &Mean values \\
\enddata
\end{deluxetable}

The IPHAS result \citep{gi11} is significantly different, and appears to imply a reddening-distance relation from IPHAS photometry for stars in this field that deviates markedly from results found here and by \citet{nk80}. As noted earlier, interstellar reddening is difficult to correlate exactly with distance in the Galactic plane, given that it appears to arise mainly in distinct dust clouds along Galactic sight lines \citep[see][]{tu94}, and that may help to explain the discrepancy in this case. Note that the discrepancy is in distance only; the color excesses found for Abell 8 agree quite closely (Table~\ref{tbl-4}).

The evidence for a physical association between Abell 8 and Bica 6 is particularly strong for the radial velocity observations, since the expected heliocentric velocity from Galactic rotation for an object 1.6 kpc distant at $\ell=167^{\circ}$ is $\sim 60$ km s$^{-1}$ more negative. It would be highly unusual to have two spatially adjacent objects sharing such unusual Galactic orbits if they were not physically associated. Certainly, according to the criteria of \citet{ma07}, Abell 8 is a likely member of Bica 6.

\section{{\rm \footnotesize EVOLUTIONARY CONSIDERATIONS}} 
\begin{figure}[!t]
  \epsscale{0.80}
  \plotone{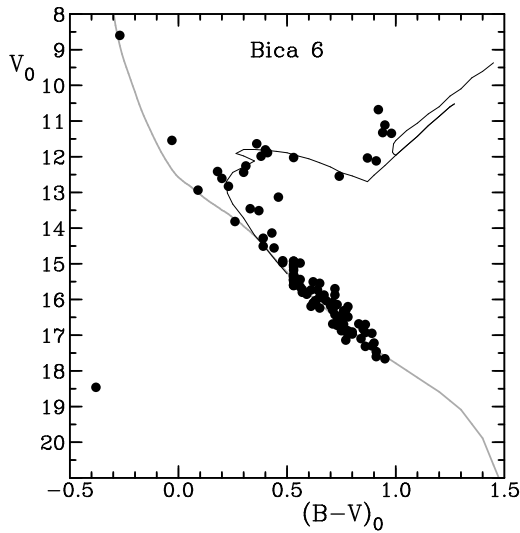}
  \caption{\small{Color-magnitude diagram for likely and potential members of Bica 6, corrected for the effects of interstellar reddening and extinction for individual stars. The gray curve represents the ZAMS for $V_0${\it --M}$_V=11.02$, the black curve is an isochrone for $\log t=9.0$. The central star of Abell 8 is the point at lower left.}}
\label{fig11}
\end{figure}

A selection of 114 stars in the Bica 6 field were identified as potential or likely cluster members according to the star counts of \citet{bo08}, the radial velocity data (Table~\ref{tbl-3}), their reddening, and inferred location in the color-magnitude diagram (Figs.~\ref{fig7}~and~\ref{fig8}). The colors and magnitudes were corrected individually for the effects of interstellar reddening and extinction, and the resulting extinction-corrected color-magnitude diagram is shown in Fig.~\ref{fig11}. As is the case with the {\it JHK}$_s$ data \citep[Fig.~\ref{fig8} and][]{bo08}, the best isochrone fit is for $\log t=9.0$.

In the evolutionary models of \citet{me93}, stars reaching the terminal stages of evolution on the main sequence for $\log t=9.0$ have masses of $\sim2.1\,M_{\sun}$, as is also the case for stars on the red giant branch of Bica 6 according to the \citet*{vd06} isochrones (not shown). The progenitor of Abell 8 (PN G167.0--00.9) must therefore have been a more massive B-type dwarf originally, perhaps up to $\sim 2.3\,M_{\sun}$ ($\sim$B9.5 V spectral type). The implied current mass of $\sim0.6\,M_{\sun}$ for the central star of Abell 8 and $\sim0.55\,M_{\sun}$ for the nebular matter \citep{ka83,ka90} fall short of the progenitor star's original mass, implying extensive mass loss during the asymptotic giant branch phase of the star to fully account for the mass budget. The excess reddening for Abell 8 found here from space reddenings for spatially adjacent stars is consistent with such a picture.

Our various photometric estimates for the brightness of the blue central star of Abell 8 (Table~\ref{tbl-5}) include eye estimates \citep[e.g.,][]{pk67} from the {\it Palomar Observatory Sky Survey} (POSS), as well as our own estimates made from image diameters measured on the POSS-II images and calibrated using similar measurements for adjacent stars of known magnitude (the present photometry). In conjunction with the field reddening of Table~\ref{tbl-4} and the derived distance to Bica 6, the data imply an absolute magnitude of $M_V = +7.44\pm0.16$ for the central star and a linear diameter for the PN of $0.47 \pm0.03$ pc, consistent with the advanced evolutionary state of the planetary nebula and the star's location towards the cool end of the PNe central star cooling sequence \citep{ka83,ka90}. For comparison purposes, a distance of 5.5 kpc for Abell 8 \citep{gi11} would imply an absolute magnitude of $M_V = +4.76$ for the central star and a linear diameter for the PN of 1.60 pc, parameters that are incompatible with each other. The former value is typical of a luminous central star that has only recently entered the PN stage \citep{be09}, while the latter value is large even for a highly evolved PN \citep{zh95}. 

Our data for cluster stars suggest that the central star of Abell 8 should have a far infrared brightness of roughly $J \simeq 20$, well beyond the magnitude cutoff of the on-line photometry for the 2MASS survey near $J = 17$, yet an image appears in the survey at $J \simeq 17.2$ at the proper location. By comparison, the star's optical brightness is near the limit of our imaging (see Figs.~\ref{fig2}~and~\ref{fig10}). It is conceivable that the infrared brightness of the object reflects emission from circumstellar dust, but additional observations are needed to assess such a possibility as well as to confirm the results. The {\it JHK}$_s$ data otherwise appear consistent with expectations for a blue central star, since the estimated {\it J--H} and {\it H--K}$_s$ colors are --0.1 and --0.1, respectively.

Although the star is single at optical wavelengths, it appears to be double in 2MASS images. The companion lies $\sim2 \arcsec$ east of the central star, and is of identical brightness in the $J$ band, being brighter at $H$ and $K_s$. Its estimated {\it J--H} and {\it H--K}$_s$ colors are +0.7 and +0.2, respectively. The companion is undetected in optical images. Both stars fit nicely into the cluster color-color and color-magnitude diagrams for 2MASS photometry (Fig.~\ref{fig8}), the companion as an M dwarf, implying that the two stars may form a wide physical binary system as well as a close optical double. The symmetrical shape of Abell 8 has been taken as evidence for a single progenitor central star by \citet{so97}, which does not necessarily contradict such arguments.

Likely and potential cluster members in Fig.~\ref{fig11} fit the $10^9$ yr isochrone reasonably well, including the diagnostic ``hook'' in the models, providing some support for the isochrones used here. Yet there are three cluster stars lying blueward of the inferred main-sequence turnoff: Star 1 (HDE~277593), B1~Vnn;, star 5, B9.5~IV; and star 12, A3~V. All three are radial velocity members, and, in the case of stars 5 and 12, closely fit the canonical description of blue stragglers. Star 1 is less easily explained, since its implied mass as a normal main-sequence star would be $\sim9-10\;M_{\sun}$, several times the turnoff point mass of the cluster. An intriguing possibility is that the star, if it is indeed a cluster member, is a post-asymptotic giant branch object caught in the act of evolving into a planetary nebula. One other example of a post-AGB object, LS~II+34$^{\circ}$~26 = V1853 Cyg \citep{pa93}, appears spectroscopically like a less-evolved object in blue-green spectra \citep{tu83,td84,tu03}, except for Balmer line emission. Star 1 may be a less extreme example of the phenomenon, given that it is not an IRAS source \citep[see][]{pa93,tu03}. The difference between its original temperature classification as B8 and its present classification as B1 may, like LS~II+34$^{\circ}$~26, be indicative of rapid evolutionary changes \citep{tu03}.

\section{{\rm \footnotesize SUMMARY}} 
Abell 8 (PN G167.0--00.9) is a very good case of a planetary nebula belonging to an open cluster, according to the spectroscopic and photometric results presented here. The inferred properties of the PN as a cluster member therefore provide useful empirical tests of the luminosities of PNe central stars as well as the PN distance scale. In the case of the latter, the implied distance of Abell 8 is consistent with the ``short'' PNe distance scale, and at variance with results implied by advocates of the ``long'' distance scale. More intriguing, however, is the cluster Bica 6 itself, since it appears to contain stars in a variety of evolutionary states associated with the evolution of main-sequence stars to their eventual fate as white dwarfs. In addition to the PN and its central star, additional likely cluster members include standard F-type and G-type giants evolved from the main sequence, two likely blue stragglers, and one star, star 1 (HDE~277593), that may be a post-AGB star evolving towards the PN state. Future additional spectroscopic and photometric observations of cluster stars would appear to be justified.

\subsection*{{\rm \scriptsize ACKNOWLEDGEMENTS}}
\small{This research used the facilities of the Canadian Astronomy Data Centre operated by the National Research Council of Canada with the support of the Canadian Space Agency. We are grateful to Peter Stetson for the use of his software routines in generating the photometric data, and to Richard Gray for providing comments on the spectral types for several cluster stars.}

%\newpage


\begin{thebibliography}{}\setlength{\itemsep}{-1.5mm}
\bibitem[Abell(1955)]{ab55} Abell, G.~O., 1955, \pasp, 67, 258
\bibitem[Abell(1966)]{ab66} Abell, G.~O., 1966, \apj, 144, 259
\bibitem[Ali(1999)]{al99} Ali, A., 1999, NewA, 4, 95
\bibitem[Acker(1978)]{ac78} Acker, A., 1978, \aaps, 33, 367
\bibitem[Acker et al.(1998)]{ac98} Acker, A., Fresneau, A., Pottasch, S.~R., \& Jasniewicz, G., 1998, \aap, 337, 253
\bibitem[Benedict et al.(2009)]{be09} Benedict, G.~F., McArthur, B.~E., Napiwotzki, R., et al., 2009, \aj, 138, 1969
\bibitem[Bianchi \& Grewing(1985)]{bg85} Bianchi, L., \& Grewing, M., 1985, IAUS, 111, 603
\bibitem[Bonatto et al.(2008)Bonatto, Bica \& Santos]{bo08} Bonatto, C., Bica, E., \& Santos, J.~F.~C.~Jr., 2008, \mnras, 386, 324
\bibitem[Cahn(1976)]{ca76} Cahn, J.~H., 1976, \aj, 81, 407
\bibitem[Cahn \& Kaler(1971)]{ca71} Cahn, J.~H., \& Kaler, J.~B., 1971, \apjs, 22, 319
\bibitem[Cahn et al.(1992)Cahn, Kaler \& Stanghellini]{ca92} Cahn, J.~H., Kaler, J.~B., \& Stanghellini, L., 1992, \aaps, 94, 399
\bibitem[Collins \& Smith(1985)]{cs85} Collins, G.~W., II, \& Smith, R.~C., 1985, \mnras, 213, 519
\bibitem[Condon \& Kaplan(1998)]{ck98} Condon, J.~J., \& Kaplan, D.~L., 1998, \apjs, 117, 361
\bibitem[Cousins \& Caldwell(2001)]{cc01} Cousins, A.~W.~J., \& Caldwell, J.~A.~R., 2001, \mnras, 323, 380
\bibitem[Cutri et al.(2003)]{cu03} Cutri, R.~M., et al., 2003, The IRSA 2MASS All-Sky Point Source Catalog of Point Sources, NASA/IPAC Infrared Science Archive
\bibitem[Dias et al.(2002)]{di02} Dias, W.~S., Alessi, B.~S., Moitinho, A., \& L\'{e}pine, J.~R.~D., 2002, \aap, 389, 871
\bibitem[Durand et al.(1998)Durand, Acker \& Zijlstra]{du98} Durand, S., Acker, A., \& Zijlstra, A., 1998, \aaps, 132, 13
\bibitem[Fernie(1963)]{fe63} Fernie, J.~D., 1963, \aj, 68, 780
\bibitem[FitzGerald(1970)]{fi70} FitzGerald, M.~P., 1970, \aap, 4, 234
\bibitem[Gathier et al.(1986)Gathier, Pottasch \& Pel]{ga86} Gathier, R., Pottasch, S.~R., \& Pel, J.~W., 1986, \aap, 157, 171
\bibitem[Giammanco et al.(2011)]{gi11} Giammanco, C., Sale, S.~E., Corradi, R.~L.~M., et al., 2011, \aap, 525, A58
\bibitem[Gray \& Corbally(2009)]{gc09} Gray, R.~O., \& Corbally, C.~J., 2009, Stellar Spectral Classification, Princeton University Press, Princeton
\bibitem[Harris et al.(1997)]{ha97} Harris, H.~C., Dahn, C.~C., Monet, D.~G., \& Pier, J.~R., 1997, in Habing H.~J., Lamers H.~J.~G.~L.~M., eds., Planetary Nebulae, IAU Symp. 180, p. 40
\bibitem[Harris et al.(2007)]{ha07} Harris, H.~C., Dahn, C.~C., Canzian, B., et al., 2007, \aj, 133, 631
\bibitem[Kaler(1983)]{ka83} Kaler, J.~B., 1983, \apj, 271, 188
\bibitem[Kaler(1985)]{ka85} Kaler, J.~B., 1985, \araa, 23, 89
\bibitem[Kaler \& Feibelman(1985)]{kf85} Kaler, J.~B., \& Feibelman, W.~A., 1985, \apj, 297, 724
\bibitem[Kaler et al.(1990)Kaler, Shaw \& Kwitter]{ka90} Kaler, J.~B., Shaw, R.~A., \& Kwitter, K.~B., 1990, \apj, 359, 392
\bibitem[Kiss et al.(2008)]{ki08} Kiss, L.~L., Szab\'{o}, G.~M., Balog Z., Parker Q.~A., \& Frew D.~J., 2008, \mnras, 391, 399
\bibitem[Kron(1978)]{kr78} Kron, G.~E., 1978, \aj, 83, 1195
\bibitem[Landolt(1983)]{la83} Landolt, A.~U., 1983, \aj, 88, 439
\bibitem[Landolt(1992)]{la92} Landolt, A.~U., 1992, \aj, 104, 340
\bibitem[Lutz(1973)]{lu73} Lutz, J.~H., 1973, \apj, 181, 135
\bibitem[Lutz(1989)]{lu89} Lutz, J.~H., 1989, in Torres-Peimbert S., ed., Planetary Nebulae, IAU Symp. 131, p. 65
\bibitem[Maciel(1984)]{ma84} Maciel, W.~J., 1984, \aaps, 55, 253
\bibitem[Majaess et al.(2007)Majaess, Turner \& Lane]{ma07} Majaess D.~J., Turner D.~G., Lane D.~J., 2007, \pasp, 119, 1349
\bibitem[Majaess et al.(2008)]{ma08} Majaess D.~J., Turner D.~G., Lane D.~J., \& Moncrieff K.~E., 2008, JAAVSO, 36, 90 
\bibitem[Meynet et al.(1993)Meynet, Mermilliod \& Maeder]{me93} Meynet G., Mermilliod J.-C., \& Maeder A., 1993, \aaps, 98, 477
\bibitem[Milne \& Aller(1975)]{ma75} Milne, D.~K., \& Aller L.~H., 1975, \aap, 38, 183
\bibitem[Moffat \& Vogt(1977)]{mv77} Moffat, A.~F.~J., \& Vogt, N., 1977, \pasp, 89, 323
\bibitem[Neckel \& Klare(1980)]{nk80} Neckel, Th., \& Klare, G., 1980, \aaps, 42, 251 
\bibitem[Osterbrock \& Ferland(2006)]{of06} Osterbrock, D.~E., \& Ferland, G.~J., 2006, Astrophysics of Gaseous Nebulae and Active Galactic Nuclei, 2nd edition, University Science Books, Sausolito, California
\bibitem[Parthasarathy(1993)]{pa93} Parthasarathy, M., 1993, \apj, 414, L109
\bibitem[Perek \& Kohoutek(1967)]{pk67} Perek, L., \& Kohoutek, L., 1967, Catalogue of Galactic Planetary Nebulae, Publication House Czechoslovak Academy of Sciences
\bibitem[Phillips(2004)]{ph04} Phillips, J.~P., 2004, \mnras, 353, 589
\bibitem[Philips et al.(2005)Philips, Cuesta \& Kemp]{ph05} Phillips, J.~P., Cuesta, L., \& Kemp, S.~N., 2005, \mnras, 357, 548
\bibitem[Pier et al.(1993)]{pi93} Pier, J.~R., Harris, H.~C., Dahn, C.~C., \& Monet, D.~G., 1993, in Weinberger R., Acker A., eds., Planetary Nebulae, IAU Symp. 155, p. 175
\bibitem[Pollacco \& Ramsay(1992)]{pr92} Pollacco, D.~L., \& Ramsay, G., 1992, \mnras, 254, 228
\bibitem[Pottasch \& Acker(1998)]{pa98} Pottasch, S.~R., \& Acker, A., 1998, \aap, 329, L5
\bibitem[Purgathofer \& Perinotto(1981)]{pp81} Purgathofer, A., \& Perinotto, M., 1981, \aap, 101, 247
\bibitem[Racine \& Harris(1975)]{rh75} Racine, R., \& Harris, W. E., 1975, \apj, 196, 413
\bibitem[Sale et al.(2009)]{sa09} Sale, S.~E., Drew, J.~E., Unruh, Y.~C., et al., 2009, \mnras, 392, 407
\bibitem[Schneider et al.(1983)]{sc83} Schneider, S.~E., Terzian, Y., Purgathofer, A., \& Perinotto, M., 1983, \apjs, 52, 399
\bibitem[Shklovsky(1956)]{sh56} Shklovsky, I.~S., 1956, \azh, 33, 222
\bibitem[Soker(1997)]{so97} Soker, N., 1997, \apjs, 112, 487
\bibitem[Terzian(1997)]{te97} Terzian, Y., 1997, in Habing H.~J., Lamers H.~J.~G.~L.~M., eds., Planetary Nebulae, IAU Symp. 180, p. 29
\bibitem[Turner(1976a)]{tu76a} Turner, D.~G., 1976a, \aj, 81, 97
\bibitem[Turner(1976b)]{tu76b} Turner, D.~G., 1976b, \aj, 81, 1125
\bibitem[Turner(1979)]{tu79} Turner, D.~G., 1979, \pasp, 91, 642
\bibitem[Turner(1980)]{tu80} Turner, D.~G., 1980, \apj, 240, 137
\bibitem[Turner(1983)]{tu83} Turner, D.~G., 1983, \aj, 86, 650
\bibitem[Turner(1992)]{tu92} Turner, D.~G., 1992, \aj, 104, 1865
\bibitem[Turner(1994)]{tu94} Turner, D.~G., 1994, \rmxaa, 29, 163
\bibitem[Turner(2003)]{tu03} Turner, D.~G. 2003, in Gray R.~O., Corbally C.~J., Philip A.~G.~D. eds., The Garrison Festschrift, L. Davis Press, Schenectady, p. 101
\bibitem[Turner(2010)]{tu10} Turner, D.~G., 2010, \apss, 326, 219
\bibitem[Turner(2011)]{tu11} Turner, D.~G., 2011, \rmxaa, 47, 127
\bibitem[Turner \& Drilling(1984)]{td84} Turner, D.~G., \& Drilling, J.~S., 1984, \pasp, 96, 292
\bibitem[Turner et al.(1986)Turner, Lyons \& Bolton]{te86} Turner, D.~G., Lyons, R.~W., \& Bolton, C.~T., 1986, Observatory, 106, 13
\bibitem[Turner et al.(2011)]{te11} Turner, D.~G., MacLellan, R.~F., Henden, A.~A., \& Berdnikov, L.~N., 2011, \rmxaa, 47, 345
\bibitem[VandenBerg et al.(2006)VandenBerg, Bergbusch \& Dowler]{vd06} VandenBerg, D.~A, Bergbusch, P.~A., \& Dowler, P.~D., 2006, \apjs, 162, 375
\bibitem[Weidemann(2000)]{we00} Weidemann, V., 2000, \aap, 363, 647
\bibitem[Zhang(1995)]{zh95} Zhang, C.~Y., 1995, \apjs, 98, 659
\bibitem[Ziznovsky(1975)]{zi75} Ziznovsky, J., 1975, BAIC, 26, 248
\end{thebibliography}
\end{document}